\documentclass[12pt]{article}

\begin{document}

\begin{center}
{\Large\bf General relativistic tidal heating for the M$\o$ller
pseudotensor}
\end{center}

\begin{center}
Lau Loi So
\end{center}


\begin{abstract}
In his study of tidal stabilization of fully relativistic neutron
stars Thorne showed that the fully relativistic expression for
tidal heating is the same as in non-relativistic Newtonian theory.
Furthermore, Thorne also noted that the tidal heating must be
independent of how one localizes gravitational energy and is
unambiguously given by that expression. Purdue and Favata
calculated the tidal heating for a number of classical
gravitational pseudotensors including that of M$\o$ller, and
obtained the result that all of them produced the same (Newtonian)
value.  However, in a re-examination of the calculation using the
M$\o$ller pseudotensor we find that there is no tidal heating.
This leads us to the conclusion that Thorne's assertion needs a
minor modification: the relativistic tidal heating is pseudotensor
independent only if the pseudotensor is derived from a Freud type
superpotential.
\end{abstract}

\section{Introduction}
Tidal heating is an empirical physical phenomenon resulting from
the net work done by an external tidal field on an isolated body.
The ocean tides on Earth provide a familiar example of this kind
of phenomenon. However, a more dramatical example is the
Jupiter-Io system, where the moon Io's active volcanoes are the
result of tidal heating~\cite{Peale}. In 1998 Thorne demonstrated
that the expected tidal heating rate is the same both in
relativistic and Newtonian gravity~\cite{Thorne}:
$\dot{W}=-\frac{1}{2}\dot{I}_{ij}E^{ij}$, where $\dot{W}$ refers
to the work rate, the dot indicates the time derivative, $I_{ij}$
is the mass quadrupole moment of the isolated body and $E_{ij}$ is
the tidal field of the external universe.  Both $I_{ij}$ and
$E_{ij}$ are time dependent, symmetric and trace free. Moreover,
Thorne also noted that this tidal heating is independent of how
one localizes the gravitational energy and is unambiguously given
by a certain value. This has been verified by calculating
$\dot{W}$ explicitly using various gravitational pseudotensors to
represent the gravitational energy-momentum density.

In 1999, Purdue used the Landau-Lifshitz pseudotensor to calculate
the tidal heating and confirmed that the result agreed with the
Newtonian perspective~\cite{Purdue,LL}. Later in 2001,
Favata~\cite{Favata} employed the same method to verify that the
Einstein, Bergmann-Thomson and M{\o}ller
pseudotensors~\cite{Freud,BT,Moller} give the same result as
Purdue found. Moreover, Booth and Creighton used the quasi-local
mass formalism of Brown and York to demonstrate the same
subject~\cite{Booth}. All of them give the same value as the
Newtonian perspective. Referring to the work of Purdue and Favata,
this seems have completed the verification that the tidal heating
is indeed independent of the gravitational pseudotensor.

Nevertheless, our re-examination of the calculation for the
M$\o$ller pseudotensor shows zero gravitational energy and
vanishing tidal heating. We suspected that Favata had used an
unsuitable extra gauge (see (73) in~\cite{Favata}) and
misinterpreted the rate of change of the constant mass
$\dot{M}\neq0$. Here we argue that obtaining the energy-momentum
pseudotensor through a Freud type superpotential guarantees the
expected tidal heating~\cite{SoarXiv} but the converse is not true
(see Sec. 3). This means that Thorne's assertion needs a minor
modification. The present paper illustrates that the relativistic
tidal heating is indeed pseudotensor independent, but only under
the condition that the pseudotensor is one of those that comes
from a superpotential which agrees with the Freud superpotential
to linear order (i.e., see (\ref{29aApril2016})).

Dirac~\cite{Dirac} mentioned that it is not possible to obtain a
gravitational field energy expression that satisfies both
conditions: (1) when added to other forms of energy the total
energy is conserved, and (2) the energy within a definite
(three-dimensional) region at a certain time is independent of the
coordinate system.  For the classical pseudotensors,  in general,
the first condition can be satisfied but the second does not. The
nice property of the M$\o$ller energy-momentum
complex~\cite{Lessner} is that the energy content of a
hypersurface does not depend on the chosen spatial coordinates,
while the complexes proposed by Einstein, Landau-Lifshitz,
Bergmann-Thomson and Goldberg do. Perhaps this may be the reason
there were many
investigators~\cite{Xulu,Yang,Radinschi,Nester1999} studying this
energy-momentum prescription in the past couple of decades. Thus
it is worthwhile to investiage the tidal heating using this
M$\o$ller pseudotensor.

\section{Technical background}
We will use $\eta_{\mu\nu}=(-1,1,1,1)$ as our spacetime
signature~\cite{MTW} and let the geometrical units $G=c=1$, where
$G$ is the Newtonian gravitational constant and $c$ the speed of
light.  We adopt the convention that Greek letters indicate
spacetime indices and Latin letters refer to spatial indices. In
principle, the classical pseudotensors~\cite{CQGSoNesterChen2009}
can be obtained from a rearrangement of the Einstein equation:
$G_{\mu\nu}=\kappa{}T_{\mu\nu}$, where the constant
$\kappa=8\pi{}G/c^{4}$ and $T_{\mu\nu}$ is the material energy
tensor. This is a basic requirement for pseudotensors (see Ch. 20
in \cite{MTW}). One can define the gravitational energy-momentum
pseudotensor in terms of a suitable superpotential
$U_{\alpha}{}^{[\mu\nu]}$:
\begin{eqnarray}
2\kappa\sqrt{-g}t_{\alpha}{}^{\mu}:=\partial_{\nu}U_{\alpha}{}^{[\mu\nu]}
-2\sqrt{-g}G_{\alpha}{}^{\mu}.\label{6aMar2015}
\end{eqnarray}
The total energy-momentum density complex can then be defined as
\begin{eqnarray}
\sqrt{-g}{\cal{T}}_{\alpha}{}^{\mu}:=\sqrt{-g}(T_{\alpha}{}^{\mu}+t_{\alpha}{}^{\mu})
=(2\kappa)^{-1}\partial_{\nu}U_{\alpha}{}^{[\mu\nu]},
\end{eqnarray}
where to get the last equality we used (\ref{6aMar2015}) and the
Einstein equation.  In vacuum it reduces to the energy
conservation relations:
$\partial_{\mu}(\sqrt{-g}t_{0}{}^{\mu})=0$. The quantity
$t_{0}{}^{0}$ and  $t_{0}{}^{j}$ can be interpreted as the
gravitational energy density and energy flux. The tidal heating
can be computed as
\begin{eqnarray}
\dot{W}=-\int_{V}\partial_{0}(\sqrt{-g}t_{0}{}^{0})d^{3}x
=\int_{V}\partial_{j}(\sqrt{-g}t_{0}{}^{j})d^{3}x.\label{21aSep2015}
\end{eqnarray}
One needs to note carefully the sign because Favata had used a
different sign for the tidal heating formula indicated in
(\ref{21aSep2015}).  As the energy-momentum can be expressed as
$P_{\mu}=(-E,\vec{P})=\int_{V}\sqrt{-g}t_{\mu}{}^{0}d^{3}x$,
Favata apparently used the wrong sign (see (58) in~\cite{Favata})
for calculating the tidal work. However Favata obtained the
correct sign for the Einstein pseudotensor simply because he had
included another negative sign for the standard Freud
superpotetial (see (17) in~\cite{Favata}). The sign for the Freud
superpotential is important and in fact it can be fixed by
evaluating the value of the ADM mass~\cite{CQGSo2009,ADM}. Using
Gauss's theorem, the last integral in (\ref{21aSep2015}) can be
converted into a surface integral of the form:
\begin{eqnarray}
\dot{W}=\oint_{\partial{}V}\sqrt{-g}t_{0}{}^{j}dS_{j},\label{14aOct2015}
\end{eqnarray}
where $dS_{j}=\hat{n}_{j}r^{2}d\Omega$,
$\hat{n}_{j}\equiv{}x_{j}/r$ is the unit radial normal vector and
$r\equiv\sqrt{\delta_{ab}x^{a}x^{b}}$ is the distance from the
body in its local asymptotic rest frame.

For the tidal heating calculation, we adopt the harmonic gauge
\begin{eqnarray}
0=\partial_{\beta}(\sqrt{-g}g^{\alpha\beta})=-\sqrt{-g}\Gamma^{\alpha\beta}{}_{\beta}.
\end{eqnarray}
This harmonic coordinate condition provides the closest
approximation to rectilinear coordinates in curved space and is
suitable for studying gravitational waves~\cite{Dirac}.   The
metric tensor can be decomposed as~\cite{Favata}
\begin{eqnarray}
g_{\mu\nu}=\eta_{\mu\nu}+\epsilon{}h_{\mu\nu}
+\epsilon^{2}k_{\mu\nu}+...,
\end{eqnarray}
where $\epsilon$ is a parameter denoting the ordering: we classify
$\eta_{\mu\nu}$ as the zeroth order, $h_{\mu\nu}$ as the 1st order
and $k_{\mu\nu}$ as the 2nd order. The traces are
$h:=\eta_{\alpha\beta}h^{\alpha\beta}$ and
$k:=\eta_{\alpha\beta}k^{\alpha\beta}$. For our tidal heating
calculation purpose, we only pay attention to the lowest
non-vanishing order; to that order we will get a relation of the
form~\cite{Purdue}
\begin{eqnarray}
\dot{W}=k_{1}\partial_{0}(I_{ij}E^{ij})+k_{2}\dot{I}_{ij}E^{ij},\label{23cSep2015}
\end{eqnarray}
where $k_{1}$, $k_{2}$ are constants. The coefficient $k_{1}$ is
related to a specific choice for the energy localization where
$\partial_{0}(I_{ij}E^{ij})$ is an ambiguous reversible
tidal-quadrupole interaction process.  We expect to get
$k_{2}=-\frac{1}{2}$ so that $-\frac{1}{2}\dot{I}_{ij}E^{ij}$ is
the unambiguous irreversible tidal heating dissipation process
that we are interested in. Therefore we only look for the tidal
heating coming from the external tidal field $E_{ij}$ interacting
with the evolving quadrupole moment $I_{ij}$ of an isolated body.
The related expressions (adopted from $(38)-(40)$
in~\cite{Purdue}) of the gravitational field tensors are:
\begin{eqnarray}
&&h_{00}=\frac{2M}{r}+\frac{3}{r^{5}}I_{ij}x^{i}x^{j}-E_{ij}x^{i}x^{j},~
h_{0j}=-\frac{2}{r^{3}}\dot{I}_{ij}x^{i}-\frac{10}{21}\dot{E}_{ik}x^{i}x^{k}x_{j}
+\frac{4}{21}\dot{E}_{ij}x^{i}r^{2}, \quad\label{27cMar2015}
\end{eqnarray}
and $h_{ij}=\delta_{ij}h_{00}$.  From (\ref{27cMar2015}), it is
easy to verify that the value of the weighting factor
$\sqrt{-g}=1+h_{00}+...$ and
\begin{eqnarray}
2h_{00,0}-\eta^{cd}h_{0c,d}=4\dot{M}r^{-1}.\label{21bDec2017}
\end{eqnarray}

According to Thorne's argument the mass $M$ is constant in
time~\cite{Thorne} and indeed the harmonic gauge at the lowest
order $\Gamma^{0\beta}{}_{\beta}=2h_{00,0}-\eta^{cd}h_{0c,d}$
demands that $\dot{M}$ is vanishing.  If the isolated body is
absorbing the external quadrupolar field, its quadrupole moment
$\dot{I}_{ij}\neq0$ in general and this can generate tidal work.
As the Einstein field equation is vanishing in vacuum, the plane
wave equation $\partial_{\lambda}\partial^{\lambda}h_{\mu\nu}=0$
appears under the criterion that the harmonic gauge is chosen.
Within our tidal heating approximation limit~\cite{Purdue},  one
should keep $\partial_{0}h_{\mu\nu}$ but can ignore
$\partial^{2}_{00}h_{\mu\nu}$ as being of higher order, so
$\vec{\nabla}^{2}h_{\mu\nu}\simeq0$ (see (51) in~\cite{Favata}).
We will come to this in Sec. 3 at (\ref{6aMay2016}).

The detailed expansion for the harmonic gauge in terms of $h$,
$hh$ and $k$ terms is
\begin{eqnarray}
\Gamma^{\alpha\beta}{}_{\beta}=\left(\epsilon-\frac{\epsilon^{2}}{2}h\right)\partial_{\beta}\bar{h}^{\alpha\beta}
+\epsilon^{2}\partial_{\beta}\left(\bar{k}^{\alpha\beta}
-h^{\alpha}{}_{\lambda}\bar{h}^{\beta\lambda}
+\frac{1}{4}\eta^{\alpha\beta}h^{\lambda\sigma}\bar{h}_{\lambda\sigma}\right),\label{6aJuly2017}
\end{eqnarray}
where
$\bar{h}^{\alpha\beta}=h^{\alpha\beta}-\frac{1}{2}\eta^{\alpha\beta}h$
and likewise
$\bar{k}^{\alpha\beta}=k^{\alpha\beta}-\frac{1}{2}\eta^{\alpha\beta}k$.
This equation gives the usual first order harmonic gauge
$\partial_{\alpha}\bar{h}^{\alpha\beta}=0$ and the second order
\begin{eqnarray}
\begin{array}{ccccc}
\partial_{\beta}(\bar{k}^{\alpha\beta}
-h^{\alpha}{}_{\lambda}\bar{h}^{\beta\lambda}
+\frac{1}{4}\eta^{\alpha\beta}h^{\lambda\sigma}\bar{h}_{\lambda\sigma})=0.\label{26aSep2017}
\end{array}
\end{eqnarray}
We split the time and spatial components for (\ref{6aJuly2017}) as
follows:
\begin{eqnarray}
\Gamma^{0\beta}{}_{\beta}&=&(2h_{00,0}-\eta^{cd}h_{0c,d})
-\frac{1}{2}\eta^{\beta\lambda}(2k_{0\beta,\lambda}-k_{\beta\lambda,0})
+h_{00}h_{00,0}-\eta^{cd}h_{0c}(h_{00,d}+h_{0d,0}),\quad\label{3aMay2016}\\
\Gamma^{j\beta}{}_{\beta}&=&\eta^{ja}\left[\frac{1}{2}\eta^{\beta\lambda}
(2k_{a\beta,\lambda}-k_{\beta\lambda,a})-h_{0a,0}
+h^{00}h_{00,a}+h^{0c}(h_{0c,a}-h_{0a,c})\right]\nonumber\\
&&+h^{j0}(h_{00,0}-\eta^{cd}h_{0c,d}).\quad\quad\label{3bMay2016}
\end{eqnarray}
Including only the terms that will contribute to the tidal heating
to the order of accuracy of our interest (\ref{23cSep2015}), which
are shown in (\ref{3aMay2016}) and (\ref{3bMay2016}), the 2nd
order harmonic gauge components of interest are
\begin{eqnarray}
2\Gamma^{0\beta}{}_{\beta}\simeq-\eta^{\beta\lambda}(2k_{0\beta,\lambda}
-k_{\beta\lambda,0})+2h^{0\beta}h_{00,\beta},\,
2\Gamma^{j\beta}{}_{\beta}\simeq\eta^{ja}\left[
\eta^{\beta\lambda}(2k_{a\beta,\lambda}-k_{\beta\lambda,a})+\partial_{a}h^{2}_{00}\right],\label{17aMay2016}
\end{eqnarray}
where $h^{2}_{00}$ means the square of $h_{00}$. Taking an
integration for $\Gamma^{j\beta}{}_{\beta}$ in (\ref{17aMay2016})
gives
\begin{eqnarray}
\int_{V}\eta^{ja}\eta^{\beta\lambda}(2k_{a\beta,\lambda}-k_{\beta\lambda,a})d^{3}x=0,\label{15aApril2017}
\end{eqnarray}
where $\oint_{\partial{}V}\eta^{ja}h^{2}_{00}dS_{a}$ is vanishing
since the integrand is an even function.  Recall the Ricci tensor
\begin{eqnarray}
R_{\alpha}{}^{\mu}=\Gamma^{\beta\nu}{}_{\alpha}\Gamma_{\beta\nu}{}^{\mu}
-\Gamma^{\beta\mu}{}_{\alpha}\Gamma_{\beta}{}^{\nu}{}_{\nu}
+\frac{1}{2}g^{\mu\rho}g^{\beta\nu}(g_{\alpha\beta,\rho\nu}+g_{\rho\beta,\alpha\nu}
-g_{\alpha\rho,\beta\nu}-g_{\beta\nu,\alpha\rho}).\label{4aFeb2016}
\end{eqnarray}
For the first order $h_{\beta\lambda}$ in vacuum, there is an
identity using (\ref{4aFeb2016}) (i.e., see (70)
in~\cite{Favata}):
\begin{eqnarray}
0=\eta^{\mu\rho}\eta^{\beta\nu}(h_{\alpha\beta,\rho\nu}+h_{\rho\beta,\alpha\nu}
-h_{\alpha\rho,\beta\nu}-h_{\beta\nu,\alpha\rho}).\label{8aApril2016}
\end{eqnarray}
Meanwhile, for the 2nd order, referring to (\ref{4aFeb2016}) again
when $(\alpha,\mu)=(0,j)$
\begin{eqnarray}
0=\eta^{ja}[3h_{00,0}h_{00,a}-\eta^{cd}h_{00,c}h_{0d,a}-h^{0b}h_{00,ab}
+\eta^{\beta\lambda}(k_{a\beta,0\lambda}+k_{0\beta,a\lambda}-k_{\beta\lambda,0a}-k_{0a,\beta\lambda})].
\label{17cMay2016}
\end{eqnarray}

Favata proposed an extra gauge (see (73) in \cite{Favata})
\begin{eqnarray}
k^{\beta}{}_{\nu,\beta}=k_{,\nu},\label{21cDec2017}
\end{eqnarray}
but we argue that this is invalid. Here we give three explanations
does not agree: (i) equation (\ref{15aApril2017}) shows that
$\int_{V}\eta^{ja}\eta^{\beta\lambda}(2k_{a\beta,\lambda}-k_{\beta\lambda})d^{3}x=0$,
and which does not agree with the Favata's gauge. (ii) apply the
harmonic and Favata gauges simultaneously in (\ref{17aMay2016}),
we expect
$\eta^{ja}\partial_{a}\Gamma^{0\beta}{}_{\beta}+\partial_{0}\Gamma^{j\beta}{}_{\beta}$
vanishes, but our calculation gives an inconsistent value
\begin{eqnarray}
\oint_{\partial{}V}(\eta^{ja}\partial_{a}\Gamma^{0\beta}{}_{\beta}
+\partial_{0}\Gamma^{j\beta}{}_{\beta})dS_{j}
=\frac{4\pi}{21}\left[\frac{48}{5}\partial_{0}(I_{ij}E^{ij})+\dot{I}_{ij}E^{ij}\right].\label{16aFeb2017}
\end{eqnarray}
Obviously, Favata's $k$ condition is incompatible with the
harmonic gauge.  (iii) we get the same non-consistent result if we
apply the harmonic gauge and the scalar Riemann curvature
$g^{\mu\nu}R_{\mu\nu}$ together in empty space:
\begin{eqnarray}
\eta^{\alpha\beta}\eta^{\mu\nu}\partial_{\nu}(k_{\alpha\beta,\mu}-k_{\alpha\mu,\beta})
=\Gamma^{\alpha\beta\nu}\Gamma_{\alpha\beta\nu},\label{21aDec2017}
\end{eqnarray}
which is not vanishing in general. Explicitly, the LHS in
(\ref{21aDec2017}) should vanishes according to Favata's gauge
indicated in (\ref{21cDec2017}), but the RHS cannot have the same
value in principle. One can double check using the tidal heating
approximation limit, in detail
\begin{eqnarray}
\eta^{\alpha\beta}\eta^{\mu\nu}\partial_{\nu}k_{\alpha\beta,\mu}
=2(2\Gamma^{\mu\nu}{}_{\alpha}\Gamma^{\alpha}{}_{\mu\nu}+\Gamma^{\alpha\beta\nu}\Gamma_{\alpha\beta\nu}),~{}
\eta^{\alpha\beta}\eta^{\mu\nu}\partial_{\nu}k_{\alpha\mu,\beta}
=4\Gamma^{\mu\nu}{}_{\alpha}\Gamma^{\alpha}{}_{\mu\nu}+\Gamma^{\alpha\beta\nu}\Gamma_{\alpha\beta\nu},
\end{eqnarray}
where
$\Gamma^{\mu\nu}{}_{\alpha}\Gamma^{\alpha}{}_{\mu\nu}=-\frac{1}{2}\eta^{c\,d}h_{00,c}h_{00,d}$
and
$\Gamma^{\alpha\beta\nu}\Gamma_{\alpha\beta\nu}=\frac{5}{2}\eta^{cd}h_{00,c}h_{00,d}$.

\section{Tidal heating from the Freud superpotential}
There are an infinite number of superpotentials, the
Freud~\cite{Freud} superpotential $_{F}U_{\alpha}{}^{[\mu\nu]}
:=-\sqrt{-g}g^{\beta\sigma}\Gamma^{\tau}{}_{\beta\lambda}
\delta^{\lambda\mu\nu}_{\tau\sigma\alpha}$ is a straightforward
expression that can be used for illustrating the tidal work.  Here
we use it reproduce the result of the tidal heating for the
Einstein pseudotensor. We have mentioned that the energy-momentum
complex can be computed as
$\sqrt{-g}{}_{E}{\cal{T}}_{\alpha}{}^{\mu}=\partial_{\nu}({}_{F}U_{\alpha}{}^{[\mu\nu]})$.
At any point, to lowest order in Riemann normal coordinates inside
matter this gives the desired energy-momentum stress tensor
$T_{\alpha}{}^{\mu}=\kappa^{-1}G_{\alpha}{}^{\mu}$~\cite{CQGSoNesterChen2009}.
In vacuum, the Einstein pseudotensor becomes
\begin{eqnarray}
2\kappa{}_{E}t_{\alpha}{}^{\mu}&=&\delta^{\mu}_{\alpha}(\Gamma^{\beta\lambda}{}_{\nu}\Gamma^{\nu}{}_{\beta\lambda}
-\Gamma^{\pi}{}_{\pi\nu}\Gamma^{\nu\lambda}{}_{\lambda})
+\Gamma^{\nu}{}_{\beta\nu}(\Gamma^{\mu\beta}{}_{\alpha}+\Gamma^{\beta\mu}{}_{\alpha})
+\Gamma^{\pi}{}_{\pi\alpha}(\Gamma^{\mu\lambda}{}_{\lambda}-\Gamma^{\lambda\mu}{}_{\lambda})\nonumber\\
&&-2\Gamma^{\beta\nu}{}_{\alpha}\Gamma^{\mu}{}_{\beta\nu}.
\end{eqnarray}
Apply the harmonic gauge, the gravitational energy density and
energy flux~\cite{Favata} are
\begin{eqnarray}
2\kappa{}_{E}t_{0}{}^{0}=-\frac{1}{2}\eta^{cd}h_{00,c}h_{00,d},\quad{}
2\kappa{}_{E}t_{0}{}^{j}=\eta^{ja}h_{00,0}h_{00,a}.\label{22aOct2015}
\end{eqnarray}
Note that the sign of the energy
$E=-\int_{V}\sqrt{-g}{}_{E}t_{0}{}^{0}d^{3}x$ is positive. Using
(\ref{14aOct2015}), we recover the known tidal work
$\dot{W}_{E}=\frac{3}{10}\partial_{0}(I_{ij}E^{ij})-\frac{1}{2}\dot{I}_{ij}E^{ij}$
as Favata obtained.

To the  order of concern here, it is sufficient to consider
superpotentials that are linear in the connection. There are only
three possible terms with suitable symmetry, one by itself is the
M$\o$ller superpotential. The general three parameter expression
can be written as
\begin{eqnarray}
U_{\alpha}{}^{[\mu\nu]}:=\sqrt{-g}(
a_{1}\delta^{\tau}_{\alpha}\Gamma^{\rho\lambda}{}_{\lambda}
+a_{2}\Gamma^{\tau\rho}{}_{\alpha}
+a_{3}\delta^{\rho}_{\alpha}\Gamma^{\lambda\tau}{}_{\lambda})\delta^{\mu\nu}_{\rho\tau},\label{29aApril2016}
\end{eqnarray}
where $a_{1}, a_{2}, a_{3}$ are real.

One limit that should be considered is the small region limit.
Around any arbitrary point, one can introduce Riemann normal
coordinates~\cite{CQGSo2009,SoNesterPRD} such that
\begin{equation}
g_{\alpha\beta}|_{0}=\eta_{\alpha\beta},\quad{}
g_{\alpha\beta,\mu}|_{0}=0,\quad{}
-3\Gamma^{\alpha}{}_{\beta\mu,\nu}|_{0}=R^{\alpha}{}_{\beta\mu\nu}+R^{\alpha}{}_{\mu\beta\nu}.
\end{equation}
According to the equivalence principle, to lowest order the
pseudotensor associated with the above superpotential should
reduce to the interior stress
\begin{eqnarray}
2\kappa{}T_{\alpha}{}^{\mu}=\frac{1}{3}\left[
(2a_{1}+3a_{2}+a_{3})R_{\alpha}{}^{\mu}-(2a_{1}+a_{3})\delta^{\mu}_{\alpha}R\right].
\end{eqnarray}
In order for this to agree with the Einstein equation, we have the
following constraints~\cite{CQGSo2009}:
\begin{eqnarray}
2a_{1}+a_{3}=3,\quad{}a_{2}=1.\label{5aMay2016}
\end{eqnarray}
Furthermore we considered the mass at null
infinity~\cite{Nester2017paper} and found
\begin{eqnarray}
\frac{1}{2}(a_{1}+a_{2})m(u)+\frac{1}{4\kappa}(a_{2}-a_{3})\frac{d}{du}\oint_{\partial{}V}c^{2}\sin\theta\,d\theta\,d\phi
\end{eqnarray}
where $c$ is the Bondi news function~\cite{Bondi}.  This gives the
Bondi mass provided that
\begin{eqnarray}
a_{1}+a_{2}=2, \quad{} a_{2}-a_{3}=0.\label{6bJuly2017}
\end{eqnarray}
Combining the results inside material and the null infinity from
(\ref{5aMay2016}) and (\ref{6bJuly2017}), we have the unique
solution for $a_{1},a_{2},a_{3}$ are all unity. This choice is the
same as that required at spatial infinity, in order to obtain the
ADM mass~\cite{MTW}.

Here we explain what we mean by a Freud type superpotential; we
decompose the Freud superpotential as follows
\begin{eqnarray}
_{F}U_{\alpha}{}^{[\mu\nu]}:=-\sqrt{-g}(\eta^{\beta\sigma}-\epsilon{}h^{\beta\sigma}+...)
\Gamma^{\tau}{}_{\beta\lambda}\delta^{\lambda\mu\nu}_{\tau\sigma\alpha}.
\label{15aOct2015}
\end{eqnarray}
Note that the linear in $\eta\Gamma$ terms give the expected
interior mass and tidal heating, while the $h\Gamma$ terms only
alter the value $\partial_{0}(I_{ij}E^{ij})$~\cite{SoarXiv}. Any
superpotential that agrees with the Freud superpotential to lowest
order in $h_{\mu\nu}:=g_{\mu\nu}-\eta_{\mu\nu}$, is referred to as
a Freud type superpotential. We know that the Landau-Lifshitz (LL)
superpotential can be identified as a Freud type superpotential
since we can raise the indices:
$_{LL}U^{\alpha[\mu\nu]}={}_{F}U_{\beta}{}^{[\mu\nu]}\sqrt{-g}g^{\alpha\beta}$,
which gives the desired interior mass and tidal work, whereas the
extra weighting factor $\sqrt{-g}$, once again, only affects
$\partial_{0}(I_{ij}E^{ij})$. There also exists another
possibility such as the Papapetrou superpotential~\cite{SoarXiv}:
\begin{eqnarray}
_{P}U^{\alpha[\mu\nu]}={}_{F}U_{\beta}{}^{[\mu\nu]}g^{\alpha\beta}
-\sqrt{-g}(g^{\rho\tau}h^{\pi\gamma}\Gamma^{\sigma}{}_{\lambda\pi}
+g^{\rho\pi}h^{\sigma\tau}\Gamma^{\gamma}{}_{\lambda\pi})
\delta_{\tau\gamma}^{\mu\nu}\delta_{\rho\sigma}^{\lambda\alpha}.
\end{eqnarray}
Referring to (\ref{29aApril2016}), to lowest order there are just
three possible superpotential terms and each term has its
characteristic features.

\subsection{The 1st term of the Freud superpotential}
When $(a_{1},a_{2},a_{3})=(1,0,0)$ referring to
(\ref{29aApril2016}), the first term of the Freud type
superpotential is
$_{1}U_{\alpha}{}^{[\mu\nu]}:=\sqrt{-g}\Gamma^{\rho\lambda}{}_{\lambda}\delta^{\mu\nu}_{\rho\alpha}$.
The corresponding contribution to the energy-momentum complex is
\begin{eqnarray}
(2\kappa){}_{1}{\cal{T}}_{\alpha}{}^{\mu}
=(\partial_{\nu}+\Gamma^{\pi}{}_{\pi\nu})\Gamma^{\rho\lambda}{}_{\lambda}\delta^{\mu\nu}_{\rho\alpha}.
\label{15aApril2016}
\end{eqnarray}
Inside matter at the origin, the energy-momentum
$\kappa{}T_{\alpha}{}^{\mu}=\frac{1}{3}(R_{\alpha}{}^{\mu}-\delta^{\mu}_{\alpha}R)$
in Riemann normal coordinates.  Upon applying the harmonic gauge
in vacuum, the pseudotensor $_{1}t_{\alpha}{}^{\mu}=0$, i.e., both
the energy density $_{1}t_{0}{}^{0}$ and energy flux
$_{1}t_{0}{}^{j}$ are zero.

\subsection{The 2nd term of the Freud superpotential: M$\o$ller's superpotential}
When $(a_{1},a_{2},a_{3})=(0,1,0)$ for (\ref{29aApril2016}), the
M$\o$ller superpotential~\cite{Moller} is recovered, more
precisely it is one-half of the magnitude, i.e.,
$_{M}U_{\alpha}{}^{[\mu\nu]}:=\sqrt{-g}\Gamma^{\lambda\sigma}{}_{\alpha}\delta^{\nu\mu}_{\lambda\sigma}$
. The associated energy-momentum complex is
\begin{eqnarray}
(2\kappa){}_{M}{\cal{}T}_{\alpha}{}^{\mu}
=2R_{\alpha}{}^{\mu}-\partial_{\alpha}\Gamma^{\mu\beta}{}_{\beta}
+g^{\beta\mu}\partial_{\alpha}\Gamma^{\nu}{}_{\beta\nu}
-2\Gamma^{\beta\nu}{}_{\alpha}\Gamma^{\mu}{}_{\beta\nu}.\label{23aMay2016}
\end{eqnarray}
Inside matter this reduces to
$2\kappa{}T_{\alpha}{}^{\mu}=R_{\alpha}{}^{\mu}$ at the origin in
Riemann normal coordinates. Since this result is not compatible
with Einstein's equation, one may have doubts that whether it is
meaningful keep calculating the tidal work? Although the M$\o$ller
pseudotensor has already failed the inside matter requirement,
this pseudotensor has the important feature that its gravitational
energy is coordinate system independent. In vacuum, using the
harmonic gauge condition, the energy density from
(\ref{23aMay2016}) is
\begin{eqnarray}
(2\kappa){}_{M}t_{0}{}^{0}
=3(h_{00,0})^{2}+h_{00}h_{00,00}+\eta^{cd}[h_{0c}h_{00,0d}-h_{00,c}h_{0d,0}
-\partial_{0}(h_{0c}h_{0d,0})]
-\frac{1}{2}\eta^{\beta\lambda}g_{\beta\lambda,00}
.\label{10aSep2015}
\end{eqnarray}
Referring to the detailed explanation from Purdue~(p.6 in
\cite{Purdue}), for a proper gravitational energy, we expect
something like $t_{0}{}^{0}\sim\eta^{cd}h_{00,c}h_{00,d}$ which
means the gravitational energy density should not involve any time
derivatives: ``This restriction has given us only products of
$\bar{h}^{\mu\nu}{}_{,\alpha}$ which will produce terms containing
the products $M^{2}, M{\cal{E}},M{\cal{I}},
{\cal{IE}},{\cal{II}},{\cal{EE}}$ for $(-g)t^{00}$ and
$M\dot{\cal{I}},M\dot{\cal{E}},{\cal{I}}\dot{\cal{E}},
{\cal{E}}\dot{\cal{I}},{\cal{I}}\dot{\cal{I}},{\cal{E}}\dot{\cal{E}}$
for $(-g)t^{0j}$.'' Thus we can immediately conclude that both
gravitational energy and tidal heating vanish to the order
considered. Explicitly, according to the accuracy limit involving
the types of terms of interest mentioned in connection with
(\ref{23cSep2015}), the tidal heating is
\begin{eqnarray}
\dot{W}_{M}=-\int_{V}\partial_{0}(\sqrt{-g}{}_{M}t_{0}{}^{0})d^{3}x=0.\label{29aOct2015}
\end{eqnarray}

The question then arises why Favata obtained the desired tidal
heating value for the M$\o$ller pseudotensor but we get null?  To
understand this discrepancy, we turn to the analysis of how Favata
obtained his expression.  Up to a sign Favata used the M$\o$ller
expression, recall the energy density for this pseudotensor
according to his gauge condition (see (74) in~\cite{Favata}):
\begin{eqnarray}
-8\pi{}_{M}\tau_{0}{}^{0}=-(1+h_{00})h_{00,00}+2(h_{00,0})^{2}
+\eta^{cd}\left[2h_{0c}h_{00,0d}-\partial_{0}(h_{0c}h_{0d,0})\right].
\end{eqnarray}
This result shows that
$\int_{V}\partial_{0}({}_{M}\tau_{0}{}^{0})d^{3}x$ can be
identified as a null tidal work since the integrand vanishes based
on the tidal heating approximation limit.  The corresponding
energy flux referring to Favata (see (75) in~\cite{Favata}) is:
\begin{eqnarray}
-8\pi{}_{M}\tau_{0}{}^{j}=\eta^{ja}[(1-3h_{00})h_{00,0a}
-2h_{00,0}h_{00,a}].\label{14bOct2015}
\end{eqnarray}
The accompanied tidal heating based on Favata (see (77)
in~\cite{Favata}) is
\begin{eqnarray}
\dot{M}_{M}
=-\oint_{\partial{}V}{}_{M}\tau_{0}{}^{j}\hat{n}_{j}r^{2}d\Omega
=-\dot{M}-\dot{I}_{ij}E^{ij}, \label{14eOct2015}
\end{eqnarray}
where $\dot{M}$ can be classified as the tidal work according to
Favata's point of view. He took the part on the RHS and combined
it with the LHS part and then divided by 2 to get the desired
result. But these $2M$ quantities are different, and the one the
RHS is, as already explained, a constant (i.e., see
(\ref{21bDec2017})). The LHS is written as $W$ in other words.
Checking $\partial_{j}(\eta^{ja}h_{00,0a})$ by using a volume
integral
\begin{eqnarray}
\frac{1}{\kappa}\frac{d}{dt}\int_{V}\vec{\nabla}^{2}h_{00}d^{3}x
=-\frac{d}{dt}\int_{V}M\delta({\vec{r}}-{\vec{r}}_{0})d^{3}x.\label{6aMay2016}
\end{eqnarray}
Favata argued that he obtained a nice value $\dot{M}$, but we
claim this is invalid. Here we have three objections for Favata's
non-vanishing tidal work. (i) he had included $\dot{M}$ and this
is prohibited by the harmonic gauge condition. (ii) We explained
that $\vec{\nabla}^{2}h_{00}$ should be vanishing based on the
plane wave equation and the tidal heating approximation limit.
(iii) one can apply the Poisson's equation to RHS in
(\ref{6aMay2016}) which is fixed at a one particular point, but
the tidal heating requires the separation between two different
neighbourhood points.

Favata had used an extra gauge
$k^{\alpha}{}_{\nu,\alpha}=k_{,\nu}$ (see (73) in~\cite{Favata})
to simplify the computation of the M$\o$ller pseudotensor, but it
is not a suitable gauge condition.  The detail is as follows: Let
\begin{eqnarray}
g^{\alpha\beta}=\eta^{\alpha\beta}+\epsilon{}h^{\alpha\beta}+\epsilon^{2}K^{\alpha\beta},\quad{}
x^{\mu'}=x^{\mu}+\epsilon\xi^{\mu}+\epsilon^{2}\chi^{\mu},
\end{eqnarray}
where $\xi^{\mu}$ and $\chi^{\mu}$ are vectors, $h^{\alpha\beta}$
and $K^{\alpha\beta}$ (i.e., some linear combination of
$k^{\alpha\beta}$ and quadratic of $h^{\alpha\beta}$) are known
functions. Consider
\begin{eqnarray}
g^{\mu'\nu'}
&=&\frac{\partial{}x^{\mu'}}{\partial{}x^{\alpha}}\frac{\partial{}x^{\nu'}}{\partial{}x^{\beta}}g^{\alpha\beta}
\nonumber\\
&=&\eta'{}^{\alpha\beta}+\epsilon{}h'^{\alpha\beta}+\epsilon^{2}K'^{\alpha\beta}+{\cal{O}}(\epsilon^{3}),
\end{eqnarray}
where
\begin{eqnarray}
h'_{\alpha}{}^{\beta}&=&h_{\alpha}{}^{\beta}+\partial_{\alpha}\xi^{\beta}+\partial^{\beta}\xi_{\alpha}\\
K'_{\alpha}{}^{\beta}&=&K_{\alpha}{}^{\beta}+\partial_{\alpha}\chi^{\beta}
+\partial^{\beta}\chi_{\alpha}
+h_{\alpha\lambda}\partial^{\lambda}\xi^{\beta}
+h^{\beta\lambda}\partial_{\lambda}\xi_{\alpha}
+(\partial_{\lambda}\xi_{\alpha})(\partial^{\lambda}\xi^{\beta}).
\end{eqnarray}
We fix the vector $\xi^{\mu}$ by the condition
$\partial_{\mu}\bar{h}^{\mu\nu}=0$, well known as the (linear
order) harmonic condition, and Favata used it. Explicitly
\begin{eqnarray}
\partial_{\beta}(h'_{\alpha}{}^{\beta}-C_{1}\delta^{\beta}_{\alpha}h')
=\partial_{\beta}(h_{\alpha}{}^{\beta}-C_{1}\delta^{\beta}_{\alpha}h)
+\partial_{\alpha}\partial_{\beta}(1-2C_{1})\xi^{\beta}
+\partial_{\beta}\partial^{\beta}\xi_{\alpha},\label{26bSep2017}
\end{eqnarray}
where the parameter $C_{1}=\frac{1}{2}$. Then (as is well known)
$\partial_{\mu}\bar{h}^{\mu\nu}=0$ leads to a wave equation for
$\xi_{\alpha}$, which has solutions for all given $h_{\mu\nu}$.
Now, how about $k^{\mu\nu}$? Favata claimed
$k^{\beta}{}_{\nu,\beta}=k_{,\nu}$. Let's consider a similar
technique for a one parameter set of conditions to fix
$\chi^{\mu}$. To wit
\begin{eqnarray}
\partial_{\beta}(K'_{\alpha}{}^{\beta}-C_{2}\delta^{\beta}_{\alpha}K')
&=&\partial_{\beta}(K_{\alpha}{}^{\beta}-C_{2}\delta^{\beta}_{\alpha}K)
+(1-2C_{2})\partial_{\alpha}\partial_{\beta}\chi^{\beta}
+\partial_{\beta}\partial^{\beta}\chi_{\alpha}\nonumber\\
&&+\partial_{\beta}\left[h_{\alpha\lambda}\partial^{\lambda}\xi^{\beta}
+h^{\beta\lambda}\partial_{\lambda}\xi_{\alpha}
+(\partial_{\lambda}\xi_{\alpha})(\partial^{\lambda}\xi^{\beta})\right]\nonumber\\
&&-C_{2}\partial_{\alpha}\left[2h^{\lambda\sigma}\partial_{\lambda}\xi_{\sigma}
+(\partial_{\lambda}\xi_{\sigma})(\partial^{\lambda}\xi^{\sigma})\right].
\end{eqnarray}
This is a 2nd order to be solved for $\chi^{\mu}$, namely
\begin{eqnarray}
(1-2C_{2})\partial_{\alpha}\partial_{\beta}\chi^{\beta}
+\partial_{\beta}\partial^{\beta}\chi_{\alpha}=\rm{known ~ terms},
\end{eqnarray}
where the RHS is made up of some known terms that are independent
of $\chi^{\mu}$. Consider the divergence of this equation
\begin{eqnarray}
2(1-C_{2})\partial^{\alpha}\partial_{\alpha}\partial_{\beta}\chi^{\beta}
=\partial_{\alpha}({\rm{known} ~ terms}).
\end{eqnarray}
Obviously that RHS is non-vanishing in general so $C_{2}=1$ is not
a viable option. However, choosing $C_{2}=\frac{1}{2}$ is
especially nice, since it gives a wave equation for
$\chi_{\alpha}$. Thus Favata's gauge condition cannot be satisfied
in general. In fact it seems that any value other than 1 for the
parameter $C_{2}$ could be used. Comparing with
(\ref{26aSep2017}), we find the 2nd order harmonic condition as
follows
\begin{eqnarray}
\partial_{\beta}\left(K_{\alpha}{}^{\beta}-\frac{1}{2}\delta^{\beta}_{\alpha}K\right)
=-\frac{1}{2}h\partial_{\beta}\bar{h}_{\alpha}{}^{\beta}+\partial_{\beta}\left(\bar{k}_{\alpha}{}^{\beta}
-h_{\alpha\lambda}\bar{h}^{\beta\lambda}
+\frac{1}{4}\delta^{\beta}_{\alpha}h^{\lambda\sigma}\bar{h}_{\lambda\sigma}\right)=0.
\end{eqnarray}

For the completeness, referring to (\ref{29aOct2015}), using the
property of the conservation of energy-momentum
$\partial_{\mu}(\sqrt{-g}t_{0}{}^{\mu})=0$. The vanishing tidal
work can be calculated from
$\oint_{\partial{}V}\sqrt{-g}{}_{M}t_{0}{}^{j}dS_{j}$, where
\begin{eqnarray}
(2\kappa){}_{M}t_{0}{}^{j}=\partial_{0}\Gamma^{\beta{}j}{}_{\beta}
+2\eta^{ja}h_{00,0}h_{00,a}.\label{29bApril2016}
\end{eqnarray}
Then we deduced
$\oint_{\partial{}V}\sqrt{-g}\partial_{0}\Gamma^{\beta{}j}{}_{\beta}dS_{j}
=-4\kappa[\frac{3}{10}\partial_{0}(I_{ij}E^{ij})-\frac{1}{2}\dot{I}_{ij}E^{ij}]$.
In particular, one can solve for the value as follows
\begin{eqnarray}
\frac{1}{2\kappa}\oint_{\partial{}V}\eta^{ja}\eta^{\beta\lambda}k_{\beta\lambda,0a}dS_{j}
=-\frac{1}{5}\partial_{0}(I_{ij}E^{ij})+\dot{I}_{ij}E^{ij}.\label{17zMay2016}
\end{eqnarray}
Meanwhile, applying the harmonic gauge, comparing
(\ref{17aMay2016}), (\ref{17cMay2016}) and (\ref{17zMay2016}), we
obtained
\begin{eqnarray}
\frac{1}{2\kappa}\oint_{\partial{}V}\eta^{ja}\eta^{\beta\lambda}k_{0a,\beta\lambda}dS_{j}
&=&\frac{3}{70}\partial_{0}(I_{ij}E^{ij})+\frac{29}{42}\dot{I}_{ij}E^{ij},\\
\frac{1}{2\kappa}\oint_{\partial{}V}\eta^{ja}\eta^{\beta\lambda}k_{0\beta,a\lambda}dS_{j}
&=&\frac{1}{7}\partial_{0}(I_{ij}E^{ij})+\frac{29}{42}\dot{I}_{ij}E^{ij},\\
\frac{1}{2\kappa}\oint_{\partial{}V}\eta^{ja}\eta^{\beta\lambda}k_{a\beta,0\lambda}dS_{j}
&=&-\frac{1}{5}\partial_{0}(I_{ij}E^{ij})+\frac{1}{2}\dot{I}_{ij}E^{ij}.
\end{eqnarray}

\subsection{The 3rd term of the Freud superpotential}
When $(a_{1},a_{2},a_{3})=(0,0,1)$ for (\ref{29aApril2016}), we
named this superpotential as $S$ and
$_{S}U_{\alpha}{}^{\mu\nu}:=\sqrt{-g}\delta^{\mu\nu}_{\alpha\tau}\Gamma^{\lambda\tau}{}_{\lambda}$.
This is the essential part which gives the desired tidal heating
expresson. The associated energy-momentum complex is
\begin{eqnarray}
(2\kappa){}_{S}{\cal{T}}_{\alpha}{}^{\mu}
=\delta^{\mu}_{\alpha}(-R+\partial_{\lambda}\Gamma^{\lambda\beta}{}_{\beta}
+\Gamma^{\beta\nu}{}_{\lambda}\Gamma^{\lambda}{}_{\beta\nu})
-(\partial_{\alpha}+\Gamma^{\pi}{}_{\pi\alpha})\Gamma^{\lambda\mu}{}_{\lambda}.
\end{eqnarray}
Inside matter,
$\kappa{}T_{\alpha}{}^{\mu}=\frac{1}{6}(R_{\alpha}{}^{\mu}-\delta^{\mu}_{\alpha}R)$
in Riemann normal coordinates. In vacuum, using the harmonic gauge
condition, this $S$ pseudotensor can be written as
\begin{eqnarray}
(2\kappa){}_{S}t_{\alpha}{}^{\mu}
=\delta^{\mu}_{\alpha}\Gamma^{\beta\nu}{}_{\lambda}\Gamma^{\lambda}{}_{\beta\nu}
-(\partial_{\alpha}+\Gamma^{\pi}{}_{\pi\alpha})\Gamma^{\lambda\mu}{}_{\lambda}.
\end{eqnarray}
The related gravitational energy density and energy flux are
\begin{eqnarray}
(2\kappa){}_{S}t_{0}{}^{0}=-\frac{1}{2}\eta^{cd}h_{00,c}h_{00,d},\quad{}
(2\kappa){}_{S}t_{0}{}^{j}=-\partial_{0}\Gamma^{\beta{}j}{}_{\beta}-\eta^{ja}h_{00,0}h_{00,a}.\label{16aAugust2017}
\end{eqnarray}
A minor difficulty for the tidal work integration for
(\ref{16aAugust2017}) is the term
$\partial_{0}\Gamma^{\beta{}j}{}_{\beta}$, but this problem can
easily be solved using (\ref{17zMay2016}). We find that not only
this energy density is the same as that of the Einstein
pseudotensor, but also it gives the tidal heating. Only this
superpotential is the essential part which contributes the desired
tidal heating.  More accurately, besides failing to meet the
inside matter result $2G^{\mu}{}_{\nu}$, we discovered that
whenever the superpotential includes this with unit magnitude, one
can guarantee that the suitable tidal heating value will be
achevied. In other words, the tidal heating is pseudotensor
dependent, i.e., not pseudotensor independent as Thorne expected
and Favata claimed to have verified~\cite{Favata}. Thorne wrote:
``Similarly, if, in our general relativistic analysis, we were to
change our energy localization by switching from the
Landau-Lifshitz pseudotensor to some other pseudotensor, or by
performing a gauge change on the gravitational field, we thereby
would alter $E_{\rm{}int}$ but leave $W$ unchanged"~(p.9 in
\cite{Thorne}). Perhaps Thorne had assumed that all pseudotensors
already had the standard form to linear order (see Ch. 20 in
\cite{MTW}).

\section{Conclusion}
Thorne argued that tidal heating is independent of how one
localizes the gravitational energy and the value is unambiguous.
Purdue and Favata used a number of well known pseudotensors to
calculate the tidal heating and verify that Thorne's assertion is
correct. However, after a re-examination of the M$\o$ller
pseudotensor, we found it gives a vanishing value, which suggests
that the tidal heating is after all pseudotensor dependent. In
coming to this conclusion, we have identified a minor revision of
Favata's calculation.  More precisely, the pseudotesor needs to
come from a superpotential that agrees with the Freud
superpotential to linear order in
$h_{\mu\nu}:=g_{\mu\nu}-\eta_{\mu\nu}$.  All of the famous
pseudotensors have this property, with the exception of the
M$\o$ller pseudotensor.


Here we emphasize that if a suitable gravitational energy-momentum
pseudotensor fulfills the Freud type superpoential condition, this
requirement ensures the expected tidal heating.  Furthermore, the
pseudotensor will not be physically satisfactory if it only
succeeds in achieving the desired tidal heating, but fails to meet
the inside matter requirement (e.g., pseudotensor S). Therefore
Thorne's assertion needs a minor modification: the relativistic
tidal heating is pseudotensor independent under the condition that
the peusdotensor is derived from a superpotential which is
linearly of the Freud type.

\section*{Acknowledgment}
The author would like to thank Dr. Peter Dobson, Professor
Emeritus, HKUST, for reading the manuscript and providing some
helpful comments.

\end{document}